\documentclass[aps,prd,letterpaper,nofootinbib,floatfix,amsmath,amssymb]{revtex4}

\usepackage{subfig}
\usepackage{graphicx}

\usepackage{xcolor} % CM: to type edits, comments in red.

\newcommand{\unit}[2][]{\text{#1}\,\text{#2}}
\newcommand{\ds}{ }
\newcommand{\diff}[1]{\text{d}#1} 
\newcommand{\abs}[1]{\left|#1\right|} % absolute value
\newcommand{\bra}[1]{\left\langle #1\right|}
\newcommand{\ket}[1]{\left|#1\right\rangle}
\newcommand{\braket}[2]{\left\langle#1|#2\right\rangle}
\newcommand{\bigO}[1]{\mathcal{O}(#1)} % big O notation

\newcommand{\vecN}[1]{\hat{#1}}

\newcommand{\trk}{\text{trk}}
\newcommand{\twr}{\text{twr}}
\newcommand{\coeff}[1]{{}\bar{#1}}

\begin{document}

\title{A formalism for constructing the QCD power spectrum with finite sampling}

\author{Mithila~Mangedarage}
\affiliation{Department of Physics, Illinois Institute of Technology, Chicago, Illinois 60616-3793, USA}

\author{Keith~Pedersen}
\affiliation{Department of Physics, Illinois Institute of Technology, Chicago, Illinois 60616-3793, USA}

\author{Zack~Sullivan}
\email{zsulliv1@illinoistech.edu}
\affiliation{Department of Physics, Illinois Institute of Technology, Chicago, Illinois 60616-3793, USA}

\date{December 24, 2025}

\begin{abstract}

Recent progress in the study of QCD phenomena with energy correlators motivates novel approaches to explore the information contained in the QCD radiation spectrum. Fox-Wolfram moments are a set of observables that characterize the angular distribution of energy flow in high-energy collisions. Contrary to their conventional application, they are a class of correlated moments that cannot be reduced to one or several characteristic ones. We present a formalism to extract their fully correlated information content, while systematically discarding small-angle sampling noise, on an event-by-event basis. We show that our approach circumvents a common misspecification of the data in terms of $\delta$-distributions, and is essential in keeping the power spectrum infrared and collinear safe. Our formalism introduces a means of accounting for the varying spatial extent of objects that enter the calculation of the power spectrum, with which experimental artifacts such as detector element geometries and measurement uncertainties can readily be incorporated.

\end{abstract}

\maketitle

%%%%%%%%%%%%%%%%%%%%%%%%%%%%%%%%%%%%%%%%%%%%%%%%%%%%%%%%%%%%%%%%%%%%%%%%
%%%%%%%%%%%%%%%%%%%%%%%%%%%%%%%%%%%%%%%%%%%%%%%%%%%%%%%%%%%%%%%%%%%%%%%%
%%%%%%%%%%%%%%%%%%%%%%%%%%%%%%%%%%%%%%%%%%%%%%%%%%%%%%%%%%%%%%%%%%%%%%%%
\section{Introduction}

While study of the QCD radiation spectrum at colliders has traditionally focused
on the study of jets
\cite{Ellis:1980wv,Ellis:1993tq,Cacciari:2008gp} and jet substructure
\cite{Larkoski_2020,Kogler_2019}, modern colliders are faced
with phenomena and challenges that motivate a return to a more continuous, global
approach to QCD events.
For example, long-distance phenomena such as the same-side ridge correlation
seen in lead ion collisions \emph{also} appears in high-multiplicity
proton
collisions~\cite{Khachatryan:2010gv, ATLAS-CONF-2015-027, ALICE:2021nir}.
Limitations from the coarse pointing resolution provided by energy deposits by neutral particles in calorimeters leads to large energy corrections \cite{Kakati:2024dun}. Furthermore, in high-pileup environments, where neutral particles from multiple vertices overlap, mis-assignments (``confusion'') degrade jet energy resolutions (even at large jet energies) \cite{CALICE:2022uwn}.
For example, the reconstruction of two \unit[40]{GeV} jets from a $W^\pm$
becomes a challenging affair when the final state is bathed in
hundreds of pileup vertices \cite{Calkins:2013ega}.
Thus the need for sophisticated pileup mitigation techniques is seen as crucial for ensuring the purity of the reconstructed final states \cite{SOYEZ20191}, and at the frontier of particle physics phenomenology \cite{PhysRevD.111.116010}. The correlation-based PUPPI algorithm \cite{Bertolini_2014} for pileup suppression is used in CMS analyses, even in the modern era of ML-based approaches \cite{Komiske:2017ubm, Vaughan:2025ijq}, but only sequentially considers local correlations.
In this paper we consider a global approach that captures correlations in the QCD power spectrum across all scales. We identify and
solve technical challenges with its use, and introduce a new
framework in which to study the power spectrum on an event-by-event
basis.

Fox and Wolfram created a framework to study the angular power
spectrum of QCD in their seminal papers~\cite{Fox:1978vu,Fox:1978vw},
which was used to explore average event shapes in $e^+e^-$ colliders
\cite{Ellis:1980nc,Ellis:1980wv,Vermaseren:1980qz}. Despite a long
history \cite{Ali:2010tw} of studies of the energy-energy correlations
(EEC) in the QCD power spectrum
\cite{Basham:1978bw,Fox:1980tz,Ali:1982ub}, previous papers have not
fully addressed a significant problem: QCD radiation can only be
observed through a \emph{finite} sampling of \emph{discrete}
particles. Discrete samples have a limited sampling frequency, and
inaccessible frequencies must be removed to prevent aliasing in an
analysis.  This effective band limit is normally set by the
limitations of the measuring device or analysis (e.g., the windowing
function in CMB analyses~\cite{White:1995}), but QCD events
are fundamentally different.  The smallest angular scale for which
there is meaningful information is determined by the nature of the
event itself. In this paper we will see that the primary factors
governing an event's angular resolution are the particle multiplicity
and general topology.

In Sec.~\ref{sec:power-spectrum} we solve two important challenges
to constructing a robust representation of the QCD power spectrum in
an event: (i)~a finite sample has an intrinsically limited angular
resolution, and (ii)~depending on the method of reconstruction,
detector objects have different resolutions (e.g., a charged track has
much better angular resolution than a calorimeter tower).  Both
challenges are solved in Sec.\ \ref{sec:shape-functions} by
introducing ``shape functions'' that impose a spatial band limit to
remove meaningless small-angle correlations.  Not only are shape
functions necessary for a complete decomposition of the QCD power
spectrum, they also guarantee collinear safety~\cite{Fox:1980tz}.  Our
framework should be useful for \emph{any} angular-correlation analysis
of particle-physics events, not just angular power spectra.  In
Section~\ref{sec:conclusions} we suggest future applications of the
power spectrum to better understand QCD, including tests of jet
substructure, hadronization models, and phenomenology at the CERN
Large Hadron Collider (LHC).

%%%%%%%%%%%%%%%%%%%%%%%%%%%%%%%%%%%%%%%%%%%%%%%%%%%%%%%%%%%%%%%%%%%%
%%%%%%%%%%%%%%%%%%%%%%%%%%%%%%%%%%%%%%%%%%%%%%%%%%%%%%%%%%%%%%%%%%%%
%%%%%%%%%%%%%%%%%%%%%%%%%%%%%%%%%%%%%%%%%%%%%%%%%%%%%%%%%%%%%%%%%%%%

\section{The angular power spectrum of QCD}\label{sec:power-spectrum}

The QCD power spectrum encodes the amount of energy correlated at various angular separations. The $\ell$-th moment of the power spectrum $H_{\ell}$ is given by
\begin{equation}\label{eq:H_l-double-integral}
	H_\ell
	\equiv
	\int_{\Omega} d\Omega
	\int_{\Omega^{\prime}} d\Omega^{\prime}
	\rho \left( \hat{r} \right) \rho \left( \hat{r}^\prime \right)
	P_\ell \left( \hat{r}  \cdot \hat{r}^\prime \right)
	\, ,
\end{equation} 
where $\rho(\hat{r})$ is a scalar function on the surface of the unit sphere $S^{2}$ that gives the energy of a particle traveling in the radial direction $\hat{r}$, normalized to the total energy of the event. The angular correlation between a point at $\hat{r} = (\theta, \phi)$ and another at $\hat{r}^\prime = (\theta^\prime, \phi^\prime)$ is embedded in the Legendre polynomial $P_{\ell}   $ through its argument as $(\hat{r}\cdot\hat{r}^\prime)$. Although evaluating angular positions $\hat{r}$ and $\hat{r}^\prime$ directly would require a specific choice of axes, the $\ell$-th moment of their correlation is manifestly rotationally invariant. $H_\ell$ captures every point's correlation to all points on the sphere, including itself.

The QCD power spectrum does not dictate a preference to a specific choice of $\rho( \hat{r})$. One choice would be a discrete energy distribution that describes a finite set of points on a sphere concentric with an $e^{+}e^{-}$ annihilation event. Each point is assigned a weight equal to the magnitude of the momentum of an out-going particle that pierces the surface at the said point, normalized to the total energy of the event \cite{Fox:1978vw}. For a finite set of $N$ massless particles this is
\begin{equation} \label{eq:energy-dist-delta}
	\rho(\hat{r}) 
	= \sum_{i=1}^{N} f_{i}
	\frac
	{\delta^{2}(\hat{r} - \hat{p}_{i})}
	{\sin\theta_i}
	\, ,
\end{equation}
where each particle travels in radial direction $\hat{p}_{i}$, and carries an energy fraction $f_{i} \equiv E_{i}/E_{tot}$.
The $\delta$-distributions that precisely isolate each particle's spatial location collapse the integrals in Eq.~\ref{eq:H_l-double-integral} to linear algebra equation 
\begin{equation}\label{eq:H_l-discrete}
	H_{\ell}     
	= 
	\sum_{i, j} f_i     P_\ell    (\hat{p}_i    \cdot   \hat{p}_j ) f_j    
	= \bra{f} P_\ell    \big(\ket{   \hat{p}} \cdot\bra{   \hat{p}}\big)\ket{f}
	\, .
\end{equation}
These are identical to the Fox-Wolfram moments (FWMs) \cite{Fox:1978vu,Fox:1978vw}
\begin{eqnarray}
	H_\ell &
	\equiv &
	\frac{4\pi}{2\ell+1}
	\sum_{m=-\ell}^{+\ell}
	\left|
	\sum_{i} Y_{\ell}^{m} \left( \Omega_{i} \right) \frac{\left| p_{i} \right|}{E_{tot}}
	\right|^{2}  \nonumber \\ 
	&=&
	\sum_{i, j} \frac{\left| p_{i} \right| \left| p_{j} \right|}{E_{tot}^{2}} \, P_{\ell}(\cos \phi_{ij})
	\, ,
        \label{eq:H_l-def_original}
\end{eqnarray}
where the indices $i, j$ run over all final state particles. In the first form of $H_{\ell}$, the argument $\Omega_{i}$ of the spherical harmonics $Y_{\ell}^{m}$ requires the specification of a set of axes since these are the angular positions of the particles on the sphere. However, the addition theorem of $Y_{\ell}^{m}$ reduces $H_{\ell}$ to the second form that exhibits its rotational invariance since the interior angle between the $i$-th and $j$-th particles $\phi_{ij}$ is rotationally invariant. 

The first few individual moments (Eqs.~\ref{eq:H_l-discrete} and \ref{eq:H_l-def_original}) of the power spectrum constructed with the discrete, point-like energy distributions (Eq.~\ref{eq:energy-dist-delta}) have seen continued use in a variety of contexts since their inception. $H_{2}$ and $H_{3}$ were proposed to discern between 2-jet and 3-jet processes \cite{Fox:1978vu}, much like the familiar global event shape variables from the early days of QCD~\cite{Das-Gupta:694105}. ($H_{2}$ is, in fact, related to the $C$-parameter~\cite{Hoang:2014wka}.) Their utility in event classification was naturally adopted for noise suppression at \emph{B} factories. The BABAR Collaboration found that applying a loose cut on $H_{2}$ removed diphoton or dilepton background substantially while minimally impacting the signal B candidate \cite{BaBar:2014omp}. At the Belle experiment, a linear combination of the first four moments, referred to as the ``Super Fox-Wolfram Moment,'' was used  to suppress continuum background \cite{Belle:2002vxx}. In the Higgs sector, FWMs are employed in global event topology analyses of multi-jet final states \cite{Bernaciak:2012nh, Bernaciak:2013dwa, ATLAS:2024pov} where the correlations \emph{between} moments is identified as a key issue to be understood. In machine learning approaches, up to the first six FWM's $H_{1}$--$H_{6}$ have been used either individually for single-moment analyses \cite{Field:1996nq}, or collectively as physics-informed training variables  for neural networks \cite{Erdmann:2018shi}, including recently, at the ATLAS experiment to train three different support vector machines \cite{ATLAS:2022opp}. 

\subsection{Correlated information in power spectra: global overview}\label{subsec:power-spectra-global-overview}

\begin{figure*}[tbp]
\subfloat[\label{fig:2-jet-unsmeared}2-jet-like]{
\includegraphics[width=0.5\textwidth]{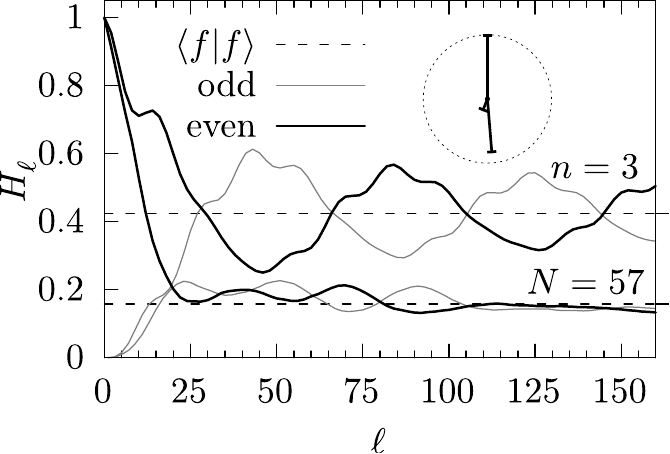}
}
\subfloat[\label{fig:3-jet-unsmeared}3-jet-like]{\includegraphics[width=0.5\textwidth]{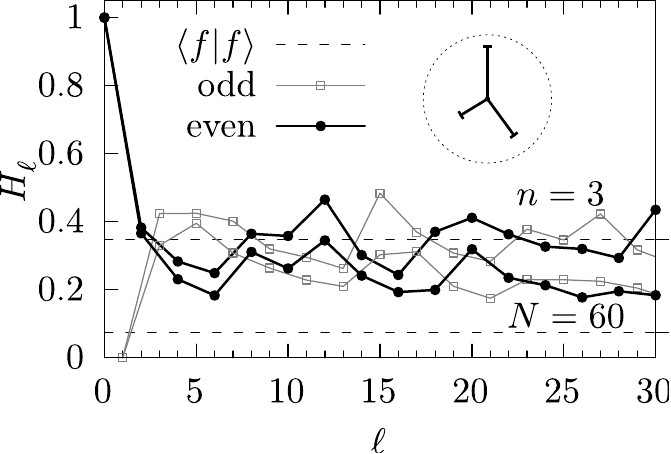}
}
\caption{The power spectra for two $e^{+}e^{-}\to q\bar{q}g$ events at $\sqrt{S}=\text{250} \, \text{GeV}$ which are
(a)~2-jet-like and (b)~3-jet-like. Lines connect even and odd moments at integer $\ell$. Each figure shows $H_\ell $ for (upper) $n=3$ initiating partons and (lower) $N=\mathcal{O}(60)$ measurable particles (after showering and hadronization). A dashed line shows the value of $\braket{f}{f}$ for each series, and the inset depicts the energy and orientation of the original partons.}
\label{fig:Hl-intro}
\end{figure*}

There is far more information in the power spectrum to be extracted than can be seen via a moment-by-moment or select-few-moments analysis. More critically, the specific choice of the discrete energy distribution in Eq.~\ref{eq:energy-dist-delta} dictates that the information content of an event is spread across infinitely many moments. This choice, which is implicit in the derivation of FWMs \cite{Fox:1978vw}, and consequently whenever they are used to study angular correlations, along with geometric features of the detector, introduces artifacts into the power spectrum. To observe these we simulate $e^{+} e^{-} \to q\bar{q}g$\footnote {\nobreak Note that $H_{2\ell} = 1$ and $H_{2\ell+1} = 0$ for the trivial QCD event  $e^{+}e^{-}\to q\bar{q}$, described by two back-to-back $\delta$-distributions \cite{Fox:1979fg}.} at $\sqrt{S}=\text{250} \, \text{GeV}$ using MadGraph~5~\cite{Alwall:2014hca} at tree-level. These events are then showered and hadronized using Pythia~8~\cite{Sjostrand:2006za, Sjostrand:2007gs}. Figure~\ref{fig:Hl-intro} shows the power spectra for two such event topologies whose energy distributions are described by Eq.~\ref{eq:energy-dist-delta}. Figures~\ref{fig:2-jet-unsmeared} and \ref{fig:3-jet-unsmeared} are the power spectra for respectively,  a two-jet-like event, and a three-jet-like event, where we connect even moments $H_{2\ell }$ (black lines) and odd moments $H_{2\ell +1}$ (gray lines) to aid the eye.  $H_\ell $ for the ${n=3}$ original partons are the upper pair of lines, and the ${N=\{57, 60\}}$ detected particles (hadrons, leptons, and photons) are the lower pair of lines. We can identify three striking properties of the power spectrum by examining these simple event topologies.

\begin{enumerate}
	\item Correlated information is spread across all moments, not just one or a few although that is indeed how they are conventionally utilized.
	\item The spectra of observed particles do not match those of their originating particles.\\
In the 2-jet-like event, $H_\ell$ for $\ell<10$ of the measurable particles do closely follow those of their originating partons, reflecting parton-jet duality \cite{Azimov:1984np}. However as $\ell$ grows, the detected spectra diverge from those of their partonic predecessors. To understand this low $\ell$-correspondence and high $\ell$-divergence between the spectra of extensive jets and infinitesimal partons, consider the angular scale $\xi_\ell$ associated with each moment $\ell$,
	\begin{equation}\label{eq:Hl-angular-scale}
		\xi_\ell=\frac{2\pi}{\ell}
		\,.
	\end{equation}
Low-$\ell$ moments have a coarse angular scale, and are not sensitive to granular features of the jet shape.  However, as $\ell$ increases, $H_\ell$ begins to probe the jets' spatial extent, thereby prompting the spectrum of observed particles to diverge from that of their originating partons.  
In the 3-jet-like event, jet substructure appears to have a stronger influence over $H_\ell$, since the measurable and 3-parton spectra diverge at a much lower $\ell$.
	\item Finally, perhaps the most critical feature is that $H_\ell$ in Fig.~\ref{fig:Hl-intro} tends to an asymptotic plateau of $\braket{f}{f}$, which drives a divergence in the total power $\sum_{\ell} (2\ell + 1) H_{\ell} \rightarrow \infty$. This is a direct consequence of computing $H_\ell$ with $\delta$-distributions; for a complete decomposition on a sphere, an energy distribution $\rho(\hat{r})$ must be square-integrable (i.e., $\int_\Omega d{\Omega}\,\rho^{2}(\hat{r}) < \infty$) --- the $\delta$-distributions in Eq.~\ref{eq:energy-dist-delta} are not. This further emphasizes that the conventional framework of FWMs spreads the information content of an event across infinitely many moments, and truncating to a few moments loses correlated information.
\end{enumerate}

%%%%%%%%%%%%%%%%%%%%%%%%%%%%%%%%%%%%%%%%%%%%%%%%%%%%%%%%%%%%%%%%%%%%%%%%
%%%%%%%%%%%%%%%%%%%%%%%%%%%%%%%%%%%%%%%%%%%%%%%%%%%%%%%%%%%%%%%%%%%%%%%%
%%%%%%%%%%%%%%%%%%%%%%%%%%%%%%%%%%%%%%%%%%%%%%%%%%%%%%%%%%%%%%%%%%%%%%%%
%%%%%%%%%%%%%%%%%%%%%%%%%%%%%%%%%%%%%%%%%%%%%%%%%%%%%%%%%%%%%%%%%%%%%%%%

\subsection{Detector artifacts embedded in the power spectrum}\label{subsec:detector-artifacts}

Consider a collision process that scatters particle energy isotropically and homogeneously. This model's energy distribution $\rho(\vecN{r}) = 1/(4\pi)$ has a featureless power spectrum, where $H_{0} = 1 \text{~and~} H_{\ell>0}\ds=0$.  Yet when we measure this process, the power spectra are
not featureless because each event has a \emph{finite}
number of particles, and is therefore a finite sampling of the isotropic distribution. The spectrum will also contain harmonic artifacts due to detector geometry.

\begin{figure*}[tb]
\subfloat[\label{fig:iso-trk}tracks]{\includegraphics[width=0.5\textwidth]{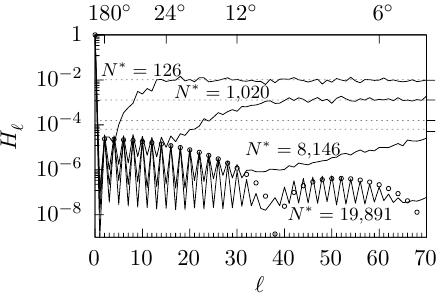}}
\subfloat[\label{fig:iso-twr}towers]{\includegraphics[width=0.5\textwidth]{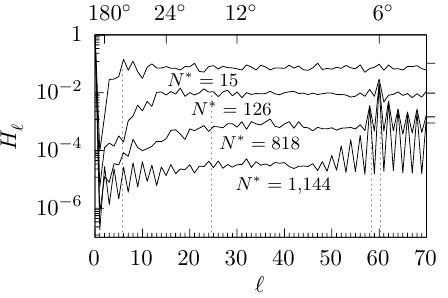}}
\caption{
The power spectrum of a random sampling of isotropically distributed particles.
Each series is labeled by the number of objects $N^*$ detected, with (a) the entire sample detected as charged tracks, and (b) the entire sample detected as neutral towers. A horizontal tick on the right edge depicts each
sample's~$\braket{f}{f}$.}
\label{fig:iso}
\end{figure*}

There are two such irreducible detector features that create angular artifacts:
(i)~the inactive beam hole, and (ii)~the finite angular precision of
calorimeter towers.  To study these effects, we construct a nearly
truth-level pseudo-detector,\footnote {A calorimeter with perfect
  energy resolution, built from towers of solid
  angle~$\Omega_\twr\ds=(6^\circ)^2$, that detects particles out to
  $|\eta_{\max}^\trk|=|\eta_{\max}^\twr|= 3$
  (the edge of the beam hole).  Since neutral particles are not
  tracked, each tower is treated as massless object with energy
  $E_\twr^0 = E_\twr\ds - \underset{\text{tracks}}{\sum_i}
  \abs{\vec{p}_i\ds}$, where the momentum from perfectly reconstructed
  massless charged particles is subtracted.}
through which we pass random isotropic samplings of $N=\lbrace16, 128,
1024, 8192, 20000\rbrace$ 
particles.  Each event is detected twice; first by assuming every
particle is charged (a track-only detection to study the beam hole
effect), then by assuming all particles are neutral (a tower-only detection
to study the effect of finite calorimeter size).  Each track or tower becomes a
$\delta$-distribution in Eq.~\ref{eq:energy-dist-delta}.  The power spectra of four such events,
with widely varying particle multiplicities~$N$, are shown in
Fig.~\ref{fig:iso} (which connect \emph{adjacent} $\ell$ with
lines).

Figure~\ref{fig:iso-trk}'s track-only power spectra confirm the
approximate isotropy of the samples; ${H_{0}=1 \text{, and }H_\ell\ds\approx0}$ for $\ell>0$ (note the
log-scale).  As particle multiplicity increases, the samples become
increasingly homogeneous, diminishing the artifacts in large $\ell$ from imperfect cancellations due to finite sampling.
However, high multiplicity uncovers a low-power oscillation in the moments due to the handful of particles which escape detection through the beam holes. 
The power spectrum of the energy distribution we are sampling can be analytically computed from a multipole definition~\cite{Fox:1978vw}:
\begin{equation}\label{eq:H_l-multipole}
	H_\ell = 
	 \frac{4\pi}{2\ell+1} \sum_{m} \left| \rho_{\ell}^{m} \right|^{2} \,,
	 \text{where } 
	 \rho_{\ell}^{m} \equiv
	\int d\Omega \rho(\hat{r}) Y_{\ell}^{m*}(\hat{r})
	\, .
\end{equation}
From this, we find the power spectrum of an isotropic energy distribution on a sphere with two antipodal holes has the raw, analytic form 
\begin{equation}\label{eq:Hl-detector-analytic}
	H_\ell =
	\begin{cases}
	\cos^{2}{\lambda} & \text{if } \ell = 0 \, ,\\
	0 & \text{if } \ell \text{ is odd, and } \\
	\left( \, \int\limits_{\cos{\lambda}}^{1} P_{\ell}(x) \, dx \right)^2 & \text{if } \ell \text{ is even} \, ,
	\end{cases}
	\, 
\end{equation}
where $\lambda$ is the polar angle between the center and the edge of the beam-hole. The circles ($\circ$) in Fig.~\ref{fig:iso-trk} represent the even moments computed for $\lambda = 0.99 \, \text{rad}$, an angular size equivalent to that of our pseudo-detector. Note that in Eq.~\ref{eq:Hl-detector-analytic} the raw zeroth moment $H_{0} < 1 \text{, for } 0 < \lambda \leq \pi/2$. The remaining power $(1 - H_{0})$ gets redistributed among all other even moments. Power spectra of samples that can cover the sphere sufficiently enough pick up these even moments, while their odd moments are contaminated by sampling noise. As multiplicity grows, the even moments approach their analytic values and the odd moments approach 0. In other words, the high-multiplicity samples probe the detector holes with increasing accuracy as multiplicity grows. The $\lesssim \bigO{10^{-5}}$ power of the beam-hole artifact is indeed quite small, so it is swamped by random sampling noise in low-multiplicity samples.

The tower-only power spectra in Fig.~\ref{fig:iso-twr} exhibit an artificial $\ell=60$ correlation, corresponding to the $6^\circ$ separation (Eq.~\ref{eq:Hl-angular-scale}) of nearest neighbor towers. Although every sample is subject to this detector feature, its angular artifact is only apparent in the high-multiplicity samples where multiple towers are activated. 

\subsection{Angular resolution and the power spectrum plateau}
\label{subsec:asymptotic-plateau}
\label{subsec:angular-resolution}

While high multiplicity sampling can identify finite size detector
artifacts in a power spectrum, there is an intrinsic limit to the
angular resolution of the measured physics process determined by the
smallest angular separations in the event. This event-specific constraint leads to an asymptotic plateau similar to those observed in Figs.~\ref{fig:Hl-intro} and \ref{fig:iso}.
Each moment $H_{\ell}$ of the power spectrum is composed of an $\ell$-independent term of \textit{self} correlations and an $\ell$-dependent term of \textit{inter-particle} correlations. This can be seen by rewriting Eq.~\ref{eq:H_l-discrete}, noting that 
the argument of the Legendre polynomials $\hat{p}_i \cdot \hat{p}_j $ is unity for $i=j$ and $P_{\ell}\left(1\right) \equiv 1 \; \forall \; \ell$,
\begin{equation}\label{eq:Hl-asym}
	H_\ell  = \underset{\text{self}}{\underbrace{ \braket{f}{f}}} + 
	\underset{\text{inter-particle}}{\underbrace{ \bra{f}
	\left(P_\ell \big(\ket{\hat{p}}\cdot     \bra{\hat{p}}\big) - \mathbb{I} \right)\ket{f}}}
	\, .
\end{equation}

With each increasing moment $\ell$, $H_{\ell}$ probes inter-particle correlations at increasingly smaller angular scales up to some $\ell_{\max}$ determined by the angular scale $\xi_{\ell_{\max}}$ of the smallest inter-particle correlation (Eq.~\ref{eq:Hl-angular-scale}). Beyond this $\ell_{\max}$ there are no finer inter-particle correlations to be resolved, the moments are dominated by random sampling noise (i.e., variations in the \emph{exact} location of each \mbox{$\delta$-distribution} relative to all others), and the inter-particle correlations begin to interfere destructively.
The result is we are left with a featureless plateau beyond $\ell_{\max}$, where 
$H_{\ell} \to \braket{f}{f}$ (shown as dashed horizontal lines in Figs.\ \ref{fig:Hl-intro} and \ref{fig:iso}). The height of this plateau is given by \cite{Pedersen2018}
\begin{equation}
\braket{f}{f} = \frac{1+a}{N}\,,
\end{equation}
where $a\ge0$, and depends on the probability distribution of the
particle's energy fraction, and $N$ is the number of samples\footnote{The expected value of $\braket{f}{f}$ for our isotropic sampling is derived to be $1.278/N$ in Ref.\ \protect{\cite{Pedersen2018}}.}.

Since the asymptotic plateau emerges from the self-interactions in the transition from useful information to noise, we turn to identifying the angular resolution 
$\xi_{\min}\ds$ by appealing to the $H_\ell\ds$ inter-particle term in
Eq.~\ref{eq:Hl-asym}.
First, we create pairs from the inter-particle
angles ${\xi_{ij}\ds\equiv\arccos(\vecN{p}_i\ds\cdot\vecN{p}_j\ds)}$
and correlation weights $w_{ij}\ds = f_i\ds f_j\ds$ (keeping only $j>i$, since the weights are symmetric). We sort the pairs on the angles $\xi_{ij}$ from smallest to largest. We
find the first $n$ angles whose collective weight is larger than
$\braket{f}{f}$ (i.e., $n$ is the first index where $\sum_{k=1}^n
2\,w_k\ds \ge \braket{f}{f}$).\footnote
{The weight $w_k\ds$ is doubled because $\xi_{ij}\ds$
  is symmetric.}
The angular resolution $\xi_{\min}\ds$ can then be
computed from the weighted geometric mean of these $n$ smallest
angles,
\begin{equation}\label{eq:angular-resolution}
	\xi_{\min}\ds = \exp\left(\frac{\sum_{k=1}^{n} w_k\ds \,\log\xi_k\ds}{\sum_{k=1}^{n} w_k\ds}\right)
	\,.
\end{equation}
In Fig.~\ref{fig:iso-twr} we determine each sample's $\xi_{\min}\ds$, and show its corresponding $\ell_{\max}\ds$ with a vertical dotted line.
These points are closely aligned with the transition to the plateau region, with the caveat that in the highest resolution samples the plateau is hidden underneath the angular artifact of the
saturated calorimeter lattice and its repeating overtones as
$\ell\to\infty$.

Despite the fact that there is no useful information to be extracted beyond some $\ell_{max}$, imposing an arbitrary cutoff such that the total power $\sum_{\ell}^{\ell_{max}}(2\ell+1)H_{l} < \infty$ is not viable. Such an imposition yields an incomplete decomposition of the energy distribution $\rho(\hat{r})$; i.e., the precisely located $\delta$-distributions in Eq.~\ref{eq:energy-dist-delta} require non-zero coefficients $\rho_{\ell}^{m}$ at all $\ell \to \infty$ values for a complete decomposition. We find that constructing \textit{continuous} energy distributions by smearing the particle energies with a ``shape function" keeps the total power $\sum_{\ell} (2\ell+1)H_{\ell}$ finite while preserving all useful information and discarding spurious, high-$\ell$ correlations. We show below how shape functions can account for the varying spatial extents of objects measured at a detector and their uncertainties. Finally, we demonstrate that shape functions keep the QCD power spectrum collinear safe.

%%%%%%%%%%%%%%%%%%%%%%%%%%%%%%%%%%%%%%%%%%%%%%%%%%%%%%%%%%%%%%%%%%%%%%%%

\section{Shape functions as low-pass filters}\label{sec:shape-functions}

To obtain a power spectrum with asymptotically vanishing moments $H_{\ell} \to 0$ as $\ell \to \infty$, we construct an event's energy flow by distributing the energy of each particle with a continuous \textit{shape function} $h_{i}(\hat{r})$ that replace the $\delta$-distributions of Eq.~\ref{eq:energy-dist-delta} with
\begin{equation}\label{eq:rho-shape}
	\rho(\hat{r}) 
		= \sum_{i=1}^N f_i \,h_i (\hat{r})
	\, .
\end{equation}
The energy distributions of the form Eq.~\ref{eq:energy-dist-delta} imply that the position of each particle is known with \textit{infinite angular resolution}. To accommodate a finite angular resolution into Eq.~\ref{eq:rho-shape}, we require an $h_{i}(\hat{r})$ that \textit{smears} the energy flow of the $i$-th particle while preserving its coarse location: i.e., a normalized angular weighting such that $h_{i}(\hat{r}) = \frac{1}{E_i} \frac{dE_{i}(\hat{r}) }{d\Omega} \, \text{sr}^{-1}$. A natural choice is a normalized, pseudo-Gaussian distribution in $\theta_{i}$, the polar angle measured from the $\hat{p_{i}}$ axis, which distributes the energy flow of each particle azimuthally symmetrically about its direction of travel and observed location $\hat{p_{i}}$ \cite{Pedersen2018}:
\begin{equation} \label{eq:pseudo-normal}
	h_{i}(\hat{r})
		= \frac{1}{2 \pi \lambda^{2} (1 - e^{-2/\lambda^2})}
		\exp \left(-\frac{(1-\hat{r} \cdot \hat{p}_i)} {\lambda^{2}} \right)
	\, ,
\end{equation}
where we have used the relation $\cos{\theta_{i}=\hat{r} \cdot \hat{p_{i}}}$.
The ``smearing angle'' $\lambda$ is analogous to the RMS width of a Gaussian. This can be readily seen 
by noting that the periodic, pseudo-Gaussian Eq.~\ref{eq:pseudo-normal} mimics a canonical Gaussian in the small angle limit $\theta_{i}$ for a sufficiently small $\lambda$,
\begin{equation} \label{eq:limit-pseudo-normal}
	\lim_{\theta_{i} \to \, 0} h_{i}(\hat{r})
	\overset{\lambda \ll1}{\approx}
		\frac{1}{2 \pi \lambda^{2}} \exp \left( -\frac{\theta_i^{2}} {2 \lambda^{2}} \right)
		\, .
\end{equation}

We can compute the amount by which to smear each particle, \textit{event by event}, by relating $\lambda$ to each sample's angular resolution $\xi_{min}$. To this end, we require that some fraction  $u\in\lbrack0,1\rbrack$ of the $i$-th particle's total energy flow through $S^{2}$ be contained within a circular cap  of solid angle $\Omega=4 \pi \sin^{2}{\left(\theta_{R}/2\right)}$ concentric with its direction of travel $\hat{p_{i}}$:
\begin{equation}\label{getlambda}
	\lambda =
	\sin\left(\frac{\theta_{R}}{2}\right)
	\sqrt{\frac{-2}{\ln[1-u(1-e^{-2/\lambda^2})]}}
	\, .
\end{equation}
Here $\theta_{R}$ is the angular radius of the cap, a fixed value of the polar angle measured from the $\hat{p_{i}}$-axis. Setting $\theta_{R}=\xi_{min}$ and starting with $\lambda=0$, this transcendental equation can be solved recursively. For example, a smearing angle $\lambda \approx 2.33^{o}$ will distribute a particle's energy flow such that $u=90 \%$ of it is contained within a $\theta_{R}=5^{0}$ circular cap concentric with the point of incidence.

\subsection{Power spectra of extensive objects}

Giving an extensive shape to the objects that enter the power spectrum calculation requires modifying the simple linear algebra of Eq.~\ref{eq:H_l-discrete}. To do so, we generalize the derivation of Ref.~\cite{Pedersen2018} to apply to an arbitrary set of azimuthally symmetric shape functions $h_{i}(\hat{r})$.
Starting with the multipole definition Eq.~\ref{eq:H_l-multipole}, and
energy distributions of the form Eq.~\ref{eq:rho-shape}, we have
\begin{equation}\label{eq:h_lm-1}
	\rho_{\ell}^{m} 
	= \sum_{i=1}^N f_i h_{(i)_{\mkern 1mu \ell}}^{ \mkern 22mu m} \,, \text{where } 
	h_{(i)_{\mkern 1mu \ell}}^{ \mkern 22mu m} \equiv \int d\Omega \; h_{i}(\hat{r})Y_{\ell}^{m*}(\hat{r})
	\, .
\end{equation}
With these, Eq.~\ref{eq:H_l-multipole} becomes
\begin{equation}\label{eq:h_lm-3}
H_{\ell}
=\sum_{i,j=1}^N f_i f_j \sum_{m=-\ell}^{\ell} \bar{h}_{(i)_{\mkern 1mu \ell}}^m \bar{h}_{(j)_{\mkern 1mu \ell}}^{m *} \, ,
\end{equation}
where the $\ell$-dependent pre-factor has been absorbed into 
\begin{equation}\label{eq:h_lm-2}
\bar{h}_{(i)_{\mkern 1mu \ell}}^{ \mkern 22mu m}
\equiv \sqrt{\frac{4 \pi}{2 \ell+1}} \; h_{(i)_{\mkern 1mu \ell}}^{ \mkern 22mu m} \,.
\end{equation}
To exploit the azimuthal symmetry of the pseudo-Gaussian Eq.~\ref{eq:pseudo-normal}, we compute the $j$-th particle's coefficients $\bar{h}_{(j)_{\mkern 1mu \ell}}^{ \mkern 22mu m*}$ in its own (primed) coordinate system that has been rotated such that the $z^{\prime}$-axis is parallel to its direction of travel $\hat{p}_j$.
Given that the set of $(2\ell+1)$ spherical harmonics of a given $\ell$ transform within themselves under rotations~\cite{Arfken:379118}, 
\begin{equation}\label{wignerd}
\bar{h}_{(j)_{\mkern 1mu \ell}}^{ \mkern 22mu m*} = \sqrt{\frac{4 \pi}{2 \ell+1}} \int d\Omega^{\prime} \; h_{j}(\hat{r}^{\prime}) \sum_{m^{\prime}} D_{mm^{\prime}}^{(\ell)}(R)Y_{\ell}^{m^{\prime}}(\hat{r}^{\prime}) \, ,
\end{equation}
where $D_{mm^{\prime}}^{(\ell)}(R)$ is the $\ell$-th Wigner \textit{D}-matrix for a rotational transformation $R: \hat{r} \mapsto \hat{r}^{\prime}$.
Now, Eq.~\ref{eq:h_lm-3} becomes
\begin{widetext}
\begin{equation}
H_{\ell}
=\frac{4 \pi}{2 \ell+1}\sum_{m, m^{\prime}} \sum_{i, j}  f_i f_j D_{mm^{\prime}}^{(\ell)}(R) \int d\Omega \; h_{i}(\hat{r})Y_{\ell}^{m*}(\hat{r}) \int d\Omega^{\prime} \; h_{j}(\hat{r}^{\prime})Y_{\ell}^{m*}(\hat{r}^{\prime})\, .
\end{equation}
\end{widetext}
Due to the azimuthal symmetry of Eq.~\ref{eq:pseudo-normal}, only the $m=m^{\prime}=0$ term survives. Thus, given that $D_{00}^{(\ell)}(R)=P_{\ell}(\hat{r}\cdot \hat{r^{\prime}})$, 
\begin{equation}\label{eq:Hl-azmth-symm}
	H_\ell = 
		\sum_{i, j =1}^{N}
		\left(f_{i},\bar{h}_{(i)_{\mkern 1mu \ell}}\right)\,
		P_\ell(\hat{p}_i\cdot\hat{p}_j)\,
		\left(f_j \,\bar{h}_{(j)_{\mkern 1mu \ell}} \right)
\end{equation}
for an event composed of particles, each with an energy flow that is azimuthally symmetric about its direction of travel. If we require that the energy flow of every particle is described by \emph{the same} shape function Eq.~\ref{eq:pseudo-normal}, this simplifies to an $\ell$-dependent pre-factor that scales the raw, power spectrum of point particles Eq.~\ref{eq:H_l-discrete}:
\begin{eqnarray}
	H_\ell &=& 
	\bar{h}_{\ell}^{2} \times
	\bra{f} P_\ell    \big(\ket{   \hat{p}} \cdot\bra{   \hat{p}}\big)\ket{f} \,, \text{where} \nonumber \\
	\bar{h}_{\ell}^{2} &=& \frac{4\pi}{2\ell+1} \left| \int d\Omega h_{(i)}(\hat{r}) Y_{\ell}^{0}(\hat{r}) \right|^{2}
	\, .
        \label{eq:Hl-azi-symm-same}
\end{eqnarray} 
This is the simplest use-case for shape functions --- calculate the sample's angular resolution $\xi_{\min}\ds$, create \emph{one} pseudo-normal shape that smears each particle using Eqs.\ \ref{eq:pseudo-normal} and \ref{getlambda}, and compute the pre-factor $\bar{h}_{\ell}^{2}$ for each
particle $i$ in the sample. $\bar{h}_{\ell}$ can be determined for any $\ell$ via the recursion \cite{Pedersen2018}
\begin{equation}\label{eq:hl-pseudo-normal}
	\coeff{h}_{\ell+1}\ds = -(2\ell+1)\lambda^2\, \coeff{h}_{\ell}\ds + \coeff{h}_{\ell-1}\ds
	\,,
\end{equation}
where $\coeff{h}_0\ds=1$ and
\begin{equation}
	\coeff{h}_1\ds = \frac{1}{\tanh(\lambda^{-2})}-\lambda^2
	\,.
\end{equation}

\begin{figure*}[htpb]
\centering
\subfloat[\label{fig:Hl-attenuation-linear}]{\includegraphics[width=0.5\textwidth]{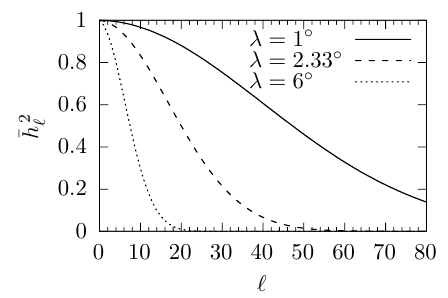}}
\subfloat[\label{fig:Hl-attenuation-log}]{\includegraphics[width=0.5\textwidth]{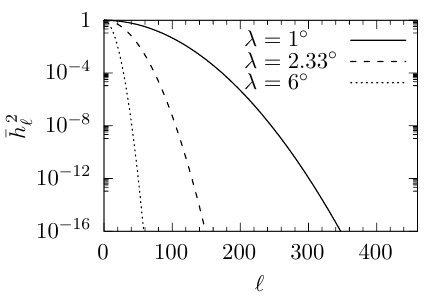}}
\caption{
	The squared coefficient $\bar{h}^2_\ell$ for the 
	pseudo-normal shape function, with several smearing angles $\lambda$
	on a (a) linear and (b) log-linear scale.}
\label{fig:Hl-attenuation}
\end{figure*}

Plotting $\coeff{h}_\ell^2$ for several values of $\lambda$ in Fig.~\ref{fig:Hl-attenuation}, we see that 
our pseudo-normal shape acts as a low-pass filter; it
preserves information at low~$\ell$, and gradually discards
information as $\ell$ increases.  As seen in
Fig.~\ref{fig:Hl-attenuation-log}, $\coeff{h}_\ell\ds$ eventually
enters an approximately exponential decay which removes $H_\ell\ds$'s
asymptotic plateau. 
Thus our shape function satisfies Parseval's relation~\cite{Varshalovich:1988ifq} 
\begin{equation}\label{eq:Parsevals_relation}
	\sum_{\ell=0}^{\infty} (2\ell+1) H_{\ell} = 4\pi \int\diff{\Omega}  \rho^{2} (\hat{r})  < \infty
	\, ,
\end{equation}
and, in effect, discards small-angle sampling noise in an under-sampled distribution to retain only the meaningful correlation structure.
A coarse-graining provided by extensive
shapes imposes a band limit through the asymptotic decay of
their~$\coeff{h}_\ell\ds$.

%%%%%%%%%%%%%%%%%%%%%%%%%%%%%%%%%%%%%%%%%%%%%%%%%%%%%%%%%%%%%%%%%%%%%%%%
\subsection{Shape functions and measurement uncertainty}\label{sec:measurement-error}

We find that shape functions provide a means of accounting for objects of varying spatial extent in an event; thereby naturally accommodating measurement uncertainty as a built-in feature. Consider, for example, a sample with a poorly measured charged track whose measurement uncertainty exceeds the sample's $\xi_{\min}\ds$. Then, its individual shape function should be replaced by its own measurement uncertainty. A more stark case is that of calorimeter towers, whose intrinsic angular uncertainty dwarfs that of tracks, and often $\xi_{\min}\ds$. (E.g., the nominal granularity of the CMS Hadron Calorimeter tower segmentation~\cite{CMS:2009gpr} is $\mathcal{O}(10^2)$ larger than the angular uncertainty of the CMS tracker~\cite{CMS:2014pgm}.)

A calorimeter is built from multiple layers of cells with varying
segmentation and sensitivity, and non-overlapping centers. This
provides detailed information about the depth and breadth of the
induced particle shower. ``Towers'' are not bins of energy with clean
edges.  While we do not attempt to simulate a realistic detector with sophisticated modeling and reconstruction algorithms that can build a probability distribution for the angular position of a
tower's initiating particle, we demonstrate a first step in how our formalism may be incorporated into such algorithms. Consider a spherical, pseudo-detector built of towers
that use a uniform \emph{circular} cap of angular radius $R$, which
provides azimuthal symmetry and approximates the correct angular correlation scale.

The procedure for measuring the angular separations of objects detected by our pseudo-detector is as follows. Given that the angular uncertainties in tracks are negligible~\cite{CMS:2014pgm} compared to those of towers~\cite{CMS:2009gpr}, we assume that angular separations $\xi_{ij}^{tt}$ between tracks can be precisely measured, even though the tracks themselves are smeared with shape functions that use their individual measurement uncertainties. Once a particle lands in a tower, however, we lose information about its precise location. In the case of a track $(i)$ that lands near a tower's centroid $(j)$, instead of the track-\emph{centroid} angle
${\xi_{ij}\ds\equiv\arccos(\vecN{p}_i\ds\cdot\vecN{p}_j\ds)}$, we use an effective angular separation between the two objects, which averages over
their constituent shape functions:
\begin{equation}\label{eq:xi_ij-extensive}
	\xi_{ij}^{\mathrm{eff}} = \int\diff{\Omega}\int\diff{\Omega^\prime}
	\,h_{(i)}\ds(\vecN{r})\,h_{(j)}\ds(\vecN{r}^\prime)\arccos(\vecN{r}\cdot\vecN{r}^\prime)
	\,.
\end{equation}
In the case of pairs of towers, we use an effective angular separation scaled by the angular radii of the two towers:
\begin{equation}\label{eq:xi_ij-Tower_Tower}
	\xi_{ij}^{\mathrm{eff}} = \frac{1}{\sqrt{R_{i}^2 + R_{j}^2}} \int\diff{\Omega}\int\diff{\Omega^\prime}
	\,h_{(i)}\ds(\vecN{r})\,h_{(j)}\ds(\vecN{r}^\prime)\arccos(\vecN{r}\cdot\vecN{r}^\prime)
	\,.
\end{equation}
Thus, with $\xi_{ij}^{tt}$ and $\xi_{ij}^{\mathrm{eff}}$ we obtain a conservative estimate for Eq.~\ref{eq:angular-resolution}, i.e., a larger $\xi_{\min}$.
By utilizing both $\xi_{ij}^{tt}$ and $\xi_{ij}^{\mathrm{eff}}$ we can capture more correlated information in an event than with a track-only scheme~\cite{STAR:2025jut}.

%%%%%%%%%%%%%%%%%%%%%%%%%%%%%%%%%%%%%%%%%%%%%%%%%%%%%%%%%%%%%%%%%%%%%%%%
%%%%%%%%%%%%%%%%%%%%%%%%%%%%%%%%%%%%%%%%%%%%%%%%%%%%%%%%%%%%%%%%%%%%%%%%
\subsection{Infrared and collinear safety}\label{sec:IRC}

Thus far in this section, we have presented a comprehensive framework of shape
functions to account for sampling granularity, the variety of detected objects that can enter the power spectrum calculation, and measurement
uncertainty.  Putting these together, for the first time we see that shape functions are essential for keeping
$H_\ell\ds$ infrared and collinear safe. 

Let us return to the two events of Fig.~\ref{fig:Hl-intro}, which were
analyzed via a truth-level detection of their measurable final-state
particles.  We now filter those particles through a pseudo-detector
(using $\Omega_{\twr}\ds=(6^\circ)^2$, 
$|\eta_{\max}^{\trk}|=|\eta_{\max}^{\twr}|=3$, and
$p_T^{\min}=\unit[300]{MeV}$), calculate the angular resolution
$\xi_{\min}\ds$ according to the procedure in Sec.~\ref{sec:measurement-error}, 
then smear the tracks via a pseudo-normal shape function with a $\lambda$ that retains $u=90\%$ of the track energy within a cap of radius $R=\xi_{\min}\ds$, to calculate
$H_\ell\ds$.  This scheme produces a smooth
event shape~$\rho(\vecN{r})$ with a sample's ``natural resolution.''
We then compute the ``angular correlation function''~\cite{Fox:1978vu,Fox:1978vw} that uses $H_\ell\ds$ as the
coefficients in a Legendre series of angular separations $\xi$,
\begin{equation}\label{eq:angular-correlation-function}
	A(\cos\xi)=\sum_{\ell=0}^\infty (2\ell+1)H_\ell\ds\,P_\ell\ds(\cos\xi)
	\,.
\end{equation}
A peak at $\cos\xi$ indicates that energy within the sample is
correlated at that angle, with the area under the peak equal to the
collective weight $w=4\sum f_i\ds f_j\ds$ of all pairs separated by
$\xi$.  For example, the large $A$ at $\cos\xi=1$ corresponds to self-correlation.

\begin{figure*}[tb]
\subfloat[\label{fig:2-jet-angular}2-jet-like]{
\includegraphics[width=0.5\textwidth]{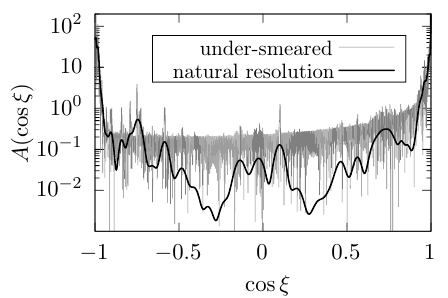}}
\subfloat[\label{fig:3-jet-angular}3-jet-like]{
\includegraphics[width=0.5\textwidth]{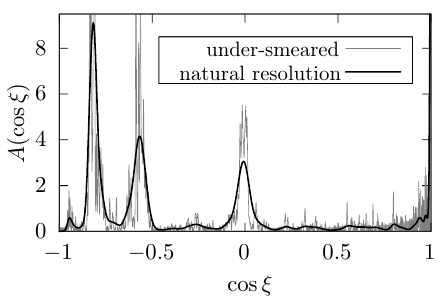}}
\caption{The angular correlation function for two 
$e^+e^-\to q\bar{q}g$ events at $\sqrt{S}=\unit[250]{GeV}$ which are
(a)~2-jet-like and (b)~3-jet-like.}
\label{fig:angular-natural-under}
\end{figure*}

It is impossible to plot $A$ for the $\delta$-function event shape used by Fox and Wolfram, since the area under the $H_\ell$ curve is infinite. However, we can
see how our pseudo-Gaussian smearing function recovers event structure by comparing to a ``nearly
$\delta$-like event shape'' by smearing every detected track or tower with a
pseudo-Gaussian ten times \emph{thinner} than the one prescribed by
the sample's angular resolution~$\xi_{\min}\ds$.  In
Fig.~\ref{fig:angular-natural-under}, we compare $A$ computed from this
``under-smeared''~$\rho(\vecN{r})$ to $A$ computed with the natural resolution shape.
The 2-jet-like event
(Fig.~\ref{fig:2-jet-angular}, shown on a log-scale) is very nearly back-to-back with peaks at $\cos\xi=\pm1$.
This is much
different than the 3-jet-like event (Fig.~\ref{fig:3-jet-angular}),
which has three large peaks (on a linear-scale) at $\cos\xi\le 0$, corresponding to the three
inter-jet angles (c.f.\ the inset of Fig.~\ref{fig:3-jet-unsmeared}).

Since each moment of $H_\ell\ds$ contains contributions from all energy correlated pairs in the event, small corrections to one energy fraction $f$ makes an infinitesimal change to the summed set of weights ${w_{ij}\ds=f_i\ds f_j\ds}$. Thus $H_\ell\ds$ (and $A$) are intrinsically
insensitive to infrared radiation.
Figure~\ref{fig:angular-natural-under} shows that shape functions
make $H_\ell\ds$ insensitive to collinear radiation (the fine
structure visible in the under-smeared $A$) as well.  This fine structure is
only probed by high-$\ell$ moments which are exponentially suppressed by our shape function. 
Hence, a low-pass filter
tailored to an event's information content provides the coarse-graining necessary to keep
the power spectrum infrared and collinear safe.  With this, we now have
a comprehensive framework in which to encode the full energy and angular
correlated information of a QCD event.

%%%%%%%%%%%%%%%%%%%%%%%%%%%%%%%%%%%%%%%%%%%%%%%%%%%%%%%%%%%%%%%%%%%%%%%%
%%%%%%%%%%%%%%%%%%%%%%%%%%%%%%%%%%%%%%%%%%%%%%%%%%%%%%%%%%%%%%%%%%%%%%%%

\section{Conclusions}\label{sec:conclusions}

In this paper we demonstrate how the effects of finite sampling and detector geometry introduce artifacts into the energy-weighted angular QCD power spectrum. We introduce a shape function to replace the precisely measured, point-particles implicit in the conventional use of FWMs~\cite{Fox:1978vw}, and suppress these artifacts. This shape function acts as a band-pass filter that smoothly removes large-moment noise. This allows us to identify the maximum available moment through which correlated information can be recovered on an event-by-event basis. We contrast this with current approaches that use only one or a combination of the first few moments. Not only do they under-utilize the information available, they abruptly cut off the power spectrum and introduce ``ringing'' artifacts into the angular correlation function Eq.~\ref{eq:angular-correlation-function} that imply false correlations at certain angular scales~\cite{Pedersen2018}.

Unlike conventional analyses with FWMs that treat all physics objects that enter a power spectrum calculation as $\delta$-functions, our formalism of shape functions has a means of accounting for the spacial extent of different objects in an event (such as tracks, towers, particle flow objects, topo-clusters, jets, etc.). Thus, shape functions naturally accommodate the measurement resolutions of tracks and calorimeter cells. While our shape function maps directly onto the leading-order shape of QCD radiation at small angles, other functions that preserve the smallest measurable angle constraint could be used to more adequately map out detector-dependent shapes unique to a particular design. We believe that current uses of FWMs, such as those at ATLAS \cite{ATLAS:2022opp, ATLAS:2024pov} and the Higgs sector \cite{Bernaciak:2012nh, Bernaciak:2013dwa}  can be improved by adopting our formalism.

By construction, the moments $H_\ell\ds$ of the power spectrum are infrared safe.  Our use of shape functions to coarse-grain an event to match its
sampled resolution guarantees collinear safety by smoothly discarding correlations at angular scales smaller than that which is determined by the information content of the detected event.  Hence, we have established a framework in which information encoded in the power spectrum can be used to extract well-defined observables using the fully-correlated data from an entire QCD event.

Attempts are already underway to incorporate FWM analyses to address potential limitations of future $e^{+}e^{-}$ colliders \cite{Li:2020vav} such as the Compact Linear Collider (CLIC) \cite{Sicking:2020gjp}. While our framework can be adopted wholesale to such efforts, the next step is to explore our framework for application at hadron colliders. Given the success of sequential algorithms such as PUPPI that consider only local correlations, we are optimistic that our global, concurrent approach to the QCD power spectrum will be robust in noisy, high-multiplicity environments. An initial foray may be found in
Ref.\ \cite{Sullivan:2019kqy}, where the formalism presented here is used to extract jet-like physics in $e^+e^-$ events with higher precision than that provided by current sequential jet algorithms in the presence of continuum noise that mimics pileup. 

Finally, our formalism to study QCD phenomena at all scales is consistent with recent efforts to modify traditional correlation methods~\cite{Alipour-fard:2024szj},  the growing desire to reformulate jet substructure within a correlation-based framework~\cite{Larkoski:2013eya, Dixon:2019uzg, Komiske:2022enw}, and on-going studies to understand short- and long-range correlations in heavy-ion collisions~\cite{Andres:2022ovj, Yang:2023dwc, Andres:2023ymw, Xing:2024yrb}. 

%%%%%%%%%%%%%%%%%%%%%%%%%%%%%%%%%%%%%%%%%%%%%%%%%%%%%%%%%%%%%%%%%%%%%%%%
%%%%%%%%%%%%%%%%%%%%%%%%%%%%%%%%%%%%%%%%%%%%%%%%%%%%%%%%%%%%%%%%%%%%%%%%
%%%%%%%%%%%%%%%%%%%%%%%%%%%%%%%%%%%%%%%%%%%%%%%%%%%%%%%%%%%%%%%%%%%%%%%%
%%%%%%%%%%%%%%%%%%%%%%%%%%%%%%%%%%%%%%%%%%%%%%%%%%%%%%%%%%%%%%%%%%%%%%%%

\begin{acknowledgments}
This paper is based upon work supported by the U.S. Department of
Energy, Office of Science, Office of High Energy Physics under Award
Number DE-SC-0008347.
\end{acknowledgments}

\bibliography{PowerSpectrum.bib}

\end{document}